\documentclass[a4paper]{jpconf}
\usepackage{graphicx}
\usepackage{amsmath,amssymb,multirow,bm}
\usepackage[colorlinks,linkcolor=blue,anchorcolor=blue,citecolor=blue,urlcolor=blue,filecolor=black]{hyperref}

\newcommand{\be}{\begin{equation}}
\newcommand{\ee}{\end{equation}}
\newcommand{\ba}{\begin{eqnarray}}
\newcommand{\ea}{\end{eqnarray}}

\begin{document}

\title{Neutron diffusion in magnetars as a source of astrophysical bursts}
\author{C. A. Bertulani$^{1}$ and R. V. Lobato$^{1, 2}$}
\address{$^{1}$ Department of Physics and Astronomy, Texas A\&M University - Commerce, Commerce, TX 75429, USA.}
\address{$^{2}$ Departamento de F\'isica, Universidad de los Andes, Bogot\'a, Colombia.}

\ead{carlos.bertulani@tamuc.edu, r.vieira@uniandes.edu.co}

\begin{abstract}
Neutron tunneling in neutron star crusts can release enormous amounts of energy on a short
timescale. We have clarified aspects of this process occurring in the outer crust regions of neutron stars when oscillations or cataclysmic events changes the crustal ambient density. We report a time-dependent Hartree-Fock-Bogoliubov model to determine the rate of neutron diffusion and conclude that a large amount of energy, in the range of $\sim 10^{40} - 10^{44}$ erg,  can be released rapidly. We suggest that this mechanism may be the source of hitherto unknown phenomena such as the  Fast Radio Bursts (FRBS). 

\end{abstract}

\section{Introduction}
{\it Isospin diffusion} is a quantitative exchange of isospin in environments with
inhomogeneous isospin content. It is a well known process which has been studied in nuclear central
collisions between heavy ions due to a {\it momentarily local imbalance} of protons and neutrons
\cite{TsangPRL.92.062701} with the goal to understand the contribution of the symmetry energy term to the equation of state of neutron stars \cite{ShiPRC.68.064604}. An extreme case of isospin imbalance
might occur in  neutron star crusts where neutron-dense regions can coexist momentarily with neutron poor
regions. In such cases, neutrons will diffuse and a homogenization of the neutron density occurs after a given time. In fact, neutron transfer in an isospin unbalanced medium has
been predicted to modify the cooling rates in accreting neutron stars
\cite{Chugunov2018}. We also infer that {\it neutron diffusion} in neutron star mergers will determine the rate to achieve local homogeneous densities. {\it Pockets, bubbles, or clumps} of
neutron-rich regions will certainly be formed as neutron stars merge or during the fallback
of a core-collapse supernovae. The diffusion mechanism will be of huge importance for nucleosynthesis because supernova fallback/accretion and neutron star mergers are the proposed sites for the
r-process~\cite{fryer/2006} responsible for the production of about half of the elements in the universe heavier than iron.  We also propose that {\it neutron diffusion by tunneling} in transient environments  can be the {\it source} of gamma-ray bursts in magnetars (see, also \cite{piro/2011, metzger/2018a}),   a source of soft gamma repeaters (SGRS) and of anomalous X-ray pulsars
(AXPs)).

Recently, the current authors have explored \cite{bertulani/2021a} the physics of isospin diffusion
in inhomogeneous neutron media due to tunneling, considering the flow of single neutrons and of
neutron-pairs passing through the potential barriers generated by the strong interaction mean-field. Neutron-rich regions  were emulated and the neutron diffusion
rates to neutron-poor regions were calculated. The goal of our study is to deduce the general properties of neutron diffusion and to identify possible scaling laws for the diffusion rates. We further suggest that the diffusion of neutrons in inhomogeneous  distributions is a trigger mechanism of short gamma/X-ray bursts/flaring activities in magnetars and in fast radio bursts (FRBs).

We have shown that significant energy release can occur by neutron diffusion in inhomogeneous neutron matter and we infer~\cite{bertulani/2021a} that neutron diffusion in such environments can be enhanced by the presence of multiple neighbors. 
Neutrons tend to flow to states with nearly the
same single-particle energies due to an effect known as {\it resonant
tunneling}. After this step, the neutrons will decay to lower energy states and also induce  nuclear many-body rearrangement of the neutron bubbles. They will also be followed by
beta-decay or gamma emission. In any of these situations, several processes can be manifested: 
(a) A quasi-instantaneous energy  energy release;  (b) Particles will be liberated in the medium composing the neutron star crust or in transient mergers; (c) Ambipolar diffusion by perturbations in huge magnetic fields can further induce non-equilibrium; (d) Electromagnetic nonlinear effects can occur in the nuclear medium. In the latter scenario, liberated particles (electrons and photons) will interact with the strong magnetic field, creating more particles (electrons and positrons) in a cascade process, that will  act as a seed for a strong electromagnetic radiation pulse.

{\it A disruptive phenomenon} within a neutron star, such as a star quake or crust fracture, will produce elastic waves that sweep across the crust, compressing and rarefying the nucleon density. The waves are followed almost instantaneously, within a time scale of $10^{-20}$s or less,  by a flurry of neutron diffusion, releasing photons and electrons. The interactions of the liberated particles with the huge magnetic field of a magnetar will produce flares responsible for the burst phenomena. The burst duration will have of the same time scale as that of the elastic waves passing through the crust.   

We have demonstrated that the loosely bound character of the transferred neutrons, together with the
presence of multiple neighbors, will lead to neutron diffusion rates that are considerably different than those obtained with popular perturbative calculations, such as the WKB approximation. The presence of multiple neighbors
leads to nonlinear effects not treatable under perturbation theory because at a given time the
decrease of the neutron density will also decrease the neutron transfer, or tunneling probability.
The neutron-neutron interaction and neutron-neutron pairing are other important physical inputs in
the calculation of the tunneling probability and diffusion rates, which are not amenable to a perturbative treatment.

\section{HFB diffusion model}
To deduce the gross properties of neutron diffusion by tunneling, we have considered a one dimensional system of neutrons. This isn't a drawback because, despite their simplicity, one-dimensional systems such as the Ising model have led to major scientific progress. In our simulations, the neutrons are immersed within a one-body potential $U(x)$
with an additional neutron-neutron interaction $v(x, x')$. We then solved the Time-Dependent Hartree-Fock-Bogoliubov (TDHFB) equations in the form
\ba
i \hbar \frac{\partial}{\partial t} u_{\alpha}(x, t)&=&\left\{-\frac{\hbar^{2} \Delta_{x}^{(2)}}{2
    m(\delta x)^{2}}+U(x)+\Gamma(x)\right\} u_{\alpha}(x, t) -\delta x \sum_{x^{\prime}}
\Delta\left(x, x^{\prime}\right) v_{\alpha}\left(x^{\prime}, t\right), \label{hfbtd1} \\
i \hbar \frac{\partial}{\partial t} v_{\alpha}(x, t)&=&-\left\{-\frac{\hbar^{2} \Delta_{x}^{(2)}}{2
    m(\delta x)^{2}}+U(x)+\Gamma^{*}(x)\right\} v_{\alpha}(x, t) - \delta x \sum_{x^{\prime}} \Delta^{*}\left(x, x^{\prime}\right) u_{\alpha}\left(x^{\prime}, t\right).  \label{hfbtd2}
\ea
The relevant quantities are defined as follows. We use the constant $\hbar^2/2m=20.73$ MeV fm$^2$, and $u_\alpha$  and $v_\alpha$ are amplitudes so that $|u_\alpha|^2$ ($|v_\alpha|^2$) is the probability that a neutron state $\alpha$ is occupied (unoccupied). The parameter $\delta x$ represents the discretization step of the one-dimensional coordinate $x$. The function $\Delta_{x}^{(2)}$ is a three-point second-order differential operator, i.e., $\Delta_x^{(2)} \phi(x)=\phi(x+\delta x)-2 \phi(x)+\phi(x-\delta x)$. Moreover,
\ba
  \Gamma(x) &=& \sum_{x^{\prime}} v\left(x-x^{\prime}\right) \rho\left(x^{\prime}, x^{\prime}\right), \\
  \Delta\left(x, x^{\prime}\right) &=& v\left(x-x^{\prime}\right) \kappa\left(x, x^{\prime}\right), \\
  \rho\left(x, x^{\prime}\right) &=& \sum_{\alpha} v_{\alpha}^{*}(x,t) v_{\alpha}\left(x^{\prime},t\right), \\ \kappa\left(x, x^{\prime}\right) &=& \sum_{\alpha} v_{\alpha}^{*}(x,t) u_{\alpha}\left(x^{\prime},t\right),
  \ea
where $\Gamma(x)$ is the interaction density, $\Delta\left(x, x^{\prime}\right)$ is the pair
correlation matrix, $\rho\left(x, x^{\prime}\right)$ is the density matrix and $\kappa\left(x, x^{\prime}\right)$ is the pairing density matrix. This set of time-dependent coupled differential equations are solved using a fourth-order Runge-Kutta algorithm.

We have obtained the initial ($t=0$) many-body wave-function by means of a diagonalisation of the  Hartree-Fock-Bogoliubov (HFB) equations based on an expansion of the single-particle wavefunctions in a harmonic oscillator basis. The required particle-number conservation was enforced using Lagrange multipliers \cite{ringschuck80}.  For a system containing $N$ neutrons, this method prepares initial states $\alpha$ and their corresponding energies and occupation numbers. A further simplification is incorporated by assuming spin symmetry, so that  $u_{\alpha\uparrow}=u_{\alpha\downarrow}$ ($v_{\alpha\uparrow}=v_{\alpha\downarrow}$), reducing the computational working space to half the number of states.

To prepare the initial states, we enforce a confining potential of the form $U_{t=0}(x) = U(x) +U_\lambda(x)$, where
\be
U(x) = {U_0\over [1+\exp\{(|x|-d)/a\}]}.
\ee
Typical values of the parameters we have used are $U_0=-100$ MeV, $d=5$ fm and $a=1$ fm. A Gaussian interaction was assumed for the particle-particle potential,
\be
v(x, x')=v_0 \exp\left(-{|x-x'|^2\over 2\sigma_0^2}\right). \label{ve}
\ee
 Here, typical parameters used are $v_0=-14$ MeV and $\sigma_0=2.5$ fm. To describe loosely-bound neutrons, mostly responsible for the tunneling process, and to simulate a chemical potential near the continuum, a confining harmonic oscillator potential is added to $U(x)$ at $t=0$, in the form  $U_\lambda (x) = \lambda x^2$.  Here, $\lambda$ is adjusted to reproduce various binding energies of the weakly-bound neutron system.

 As a prototypical system we consider clumps of $N=20$, 40, 80 and 160 neutrons. Harmonic oscillator strengths of $\lambda = 2$, $10^{-1}$, $1.15\times 10^{-2}$, and $2
 \times 10^{-3}$ MeV/fm$^{-2}$ were assumed, with the  static HFB
 equations yielding loosely bound neutrons with separation energies of $S_n=1.55,$ 0.96, 0.40,
 and 0.25 MeV for $N=20 $, 40, 80 and 160, respectively. The solution of the static HFB equations yield much larger binding energies per neutron than in a typical nucleus. But our goal is to understand the physics involved in the diffusion rates, which can be clarified  well with these values, as we are going to prove.

\section{Time evolution and diffusion rates}
After the $N$-neutron system initial wave-function is prepared, the
Eqs. \eqref{hfbtd1}-\eqref{hfbtd2} are solved by suddenly removing the confining potential
$U_\lambda$. Moreover, the single-pocket potential $U(x)$, is replaced at $t>0$ by a sequence of equally spaced Woods-Saxon-type potentials with a separation distance $D$, i.e.,
\be
U(x,d,a) \rightarrow \sum_{n=-M}^M U(x-nD,d,a).
\label{Un}
\ee
The initial $N$-neutron ssytem was placed at the center of the mean-field potential chain. The neutrons, in particular the most weakly bound ones,  tunnel through the potential barriers having an approximate width of $D-2(d+a)$. The solution of Eqs. \eqref{hfbtd1}-\eqref{hfbtd2} are limited to a box of size $L=200$ fm. Furthermore, absorbing boundary conditions were used by enforcing an imaginary potential with thickness $d_{im}= 50$ fm and strength $W_{im} = -200$ MeV at the left and right edges of the box. We allow $2M$ ($M$ an integer) of potential wells in Eq. \eqref{Un}, with $M$ depending on the separation distance $D$. For the cases considered here, we use $M$ in the range $M=2-5$ when $D$ is taken within the values of $D=20-50$ fm.

\begin{figure}
\begin{center}
\includegraphics[scale=0.45]{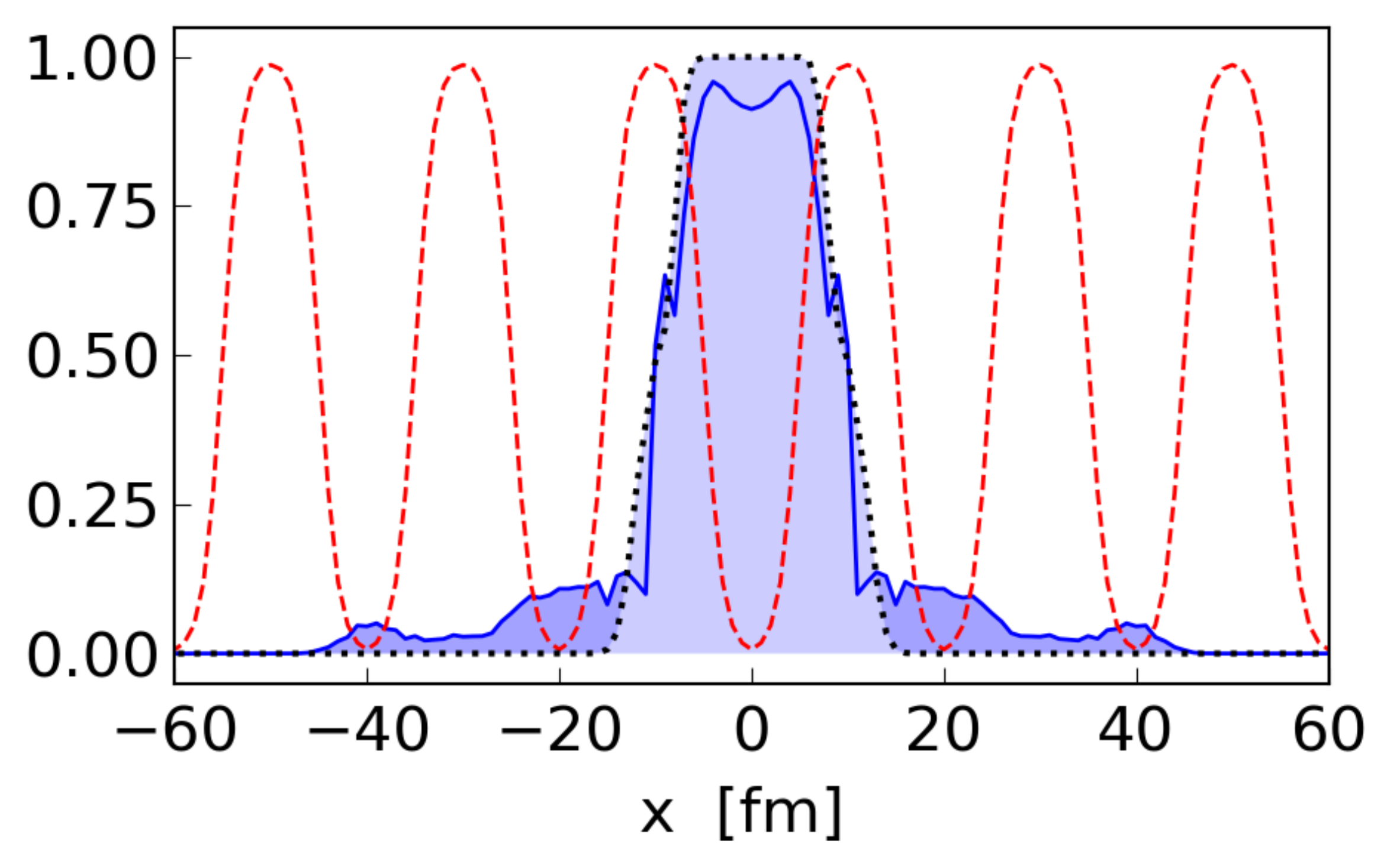}
\caption{The dashed line represents the initial, $t=0$, neutron number density for $N=160$ located at the center of a row of Woods-Saxon potentials separated by a common distance $D=20$ fm. The initial density is obtained from a HFB calculation with potential parameters described in the text. The density at $t=20,000$ fm/c is shown by the solid line, as calculated with a TDHFB method. For better visualization, the neutron number densities and Woods-Saxon potentials are multiplied by scaling factors.}
\label{diffus}
\end{center}
\end{figure}

By solving the TDHFB equations the initial N-neutron wave-function is allowed to evolve and the
transfer of neutrons from the densest regions to the neighboring, less dense, regions is obtained
as a function of time. This is displayed in Figure \ref{diffus} for the neutron number density
$\rho(x,t)$ when $N=160$ and $t=20,000$ fm/c. The initial number density is shown by a dotted line
and a light shaded area, whereas the density at time $t$ is shown by a solid curve. The nuclear
potential is indicated by dashed lines and the separation between them is taken as $D=20$ fm. The
number densities and potentials are multiplied by scaling factors for better visualization in the
same graph. The darker shaded areas in the neutron number density at time $t$ represent the leakage
of neutrons from the central potential to the neighboring ones.

The neutron density time evolution allows us to calculate the rate with which the neutrons
diffuse to neighboring regions, as well as the average diffusion coefficient ${\cal D}$ appearing in the diffusion equation
$\partial \left< \rho\right>/\partial t = {\cal D}\partial^2 \left< \rho \right>/\partial x^2$. But using the diffusion equation to infer the
neutron diffusion/tunneling rates takes a prohibitive numerical computation time. We use instead a simpler method, by monitoring the time-dependence of the number of neutrons within the initial region, to assess the diffusion rate to less dense regions. We can estimate the neutron transfer rate $\Lambda_n (t)$ by using the relation $ \Lambda_n(t) = - {dN / dt}$, with $N(t)$ being the number of neutrons within the initial confining region centered at $x=0$. It corresponds to the term with $n=0$ in Eq. \eqref{Un}  (see Fig. \ref{diffus}).

The tunneling rate according to Gamow's model \cite{Gamow1928}, widely used in the literature to estimate nuclear decay times, e.g., alpha-decay times,  is given by
\be
 \Lambda_G(t) = \nu T \sim {v \over d} \exp \left( -2 \int |\kappa (x)| dx \right). \label{Gamow}
 \ee
Here, $\nu \sim v/d$ is known as the {\it barrier assault frequency}. The local momentum of a neutron with kinetic energy $E$ is given by $\hbar \kappa = \sqrt{2m[U(x)-E]}$. The integration is carried out between the limiting  points where $E=U(x)$. Assuming that only the most energetic neutrons tunnel during a short time, the Gamow model yields increasing tunneling rates in the range $\Lambda_G \sim (0.5-0.9) \times 10^{-2}$ c/fm as the separation energies decrease within $S_n = (1.55 - 0.25)$ MeV.

\begin{figure}
\begin{center}
\includegraphics[scale=0.35]{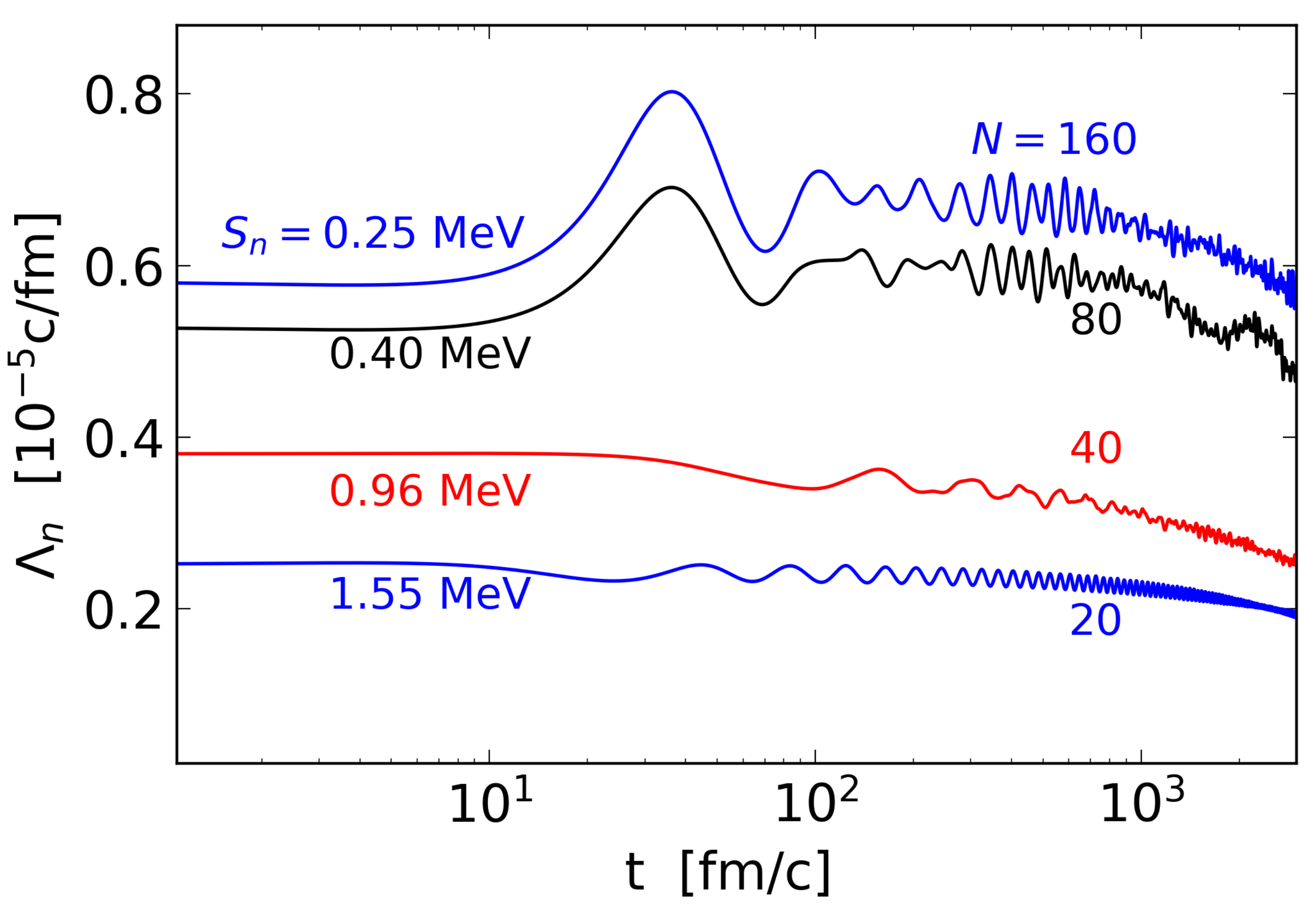}
\caption{Neutron diffusion rates as a function of time for neutron clumps immersed in a row of mean field potentials depicted in Figure \ref{diffus}.}
\label{diffus2}
\end{center}
\end{figure}

As seen in Figure \eqref{diffus2}, the neutron transfer rates $\Lambda_n$ calculated with the
dynamical microscopic method yield tunneling rates that oscillate and decrease as  time evolves. During a transient time, the rates remain
nearly constant, mainly because only the most energetic neutrons are able to leak to neighboring sites. But notably, oscillations are due to set
in a later stage simply because the system displays wave mechanical features such as reflection and interference. Then, at larger time scales, the rates drop again before they flatten out  when less-energetic neutrons begin to tunnel appreciably. This pattern likely repeats as more and more deeply bound neutrons get involved in the diffusion process. As expected, the neutron transfer rates become larger when the neutron separation energy $S_n$ is smaller.  As we show later, the rates calculated microscopically are smaller by orders of magnitude than those calculated using, e.g., a WKB transmission model.

The reason why the microscopically calculated transmission rates from the TDHFB model are smaller than
expected from a WKB approximation is partially because the two-body interaction is responsible for a larger total internal energy, $E_{int}$,  of the system. The neutron clump, or bubble, prefers to stick together due to the strong interaction and, therefore, density homogenization by tunneling is suppressed.

\begin{figure}
\begin{center}
\includegraphics[scale=0.4]{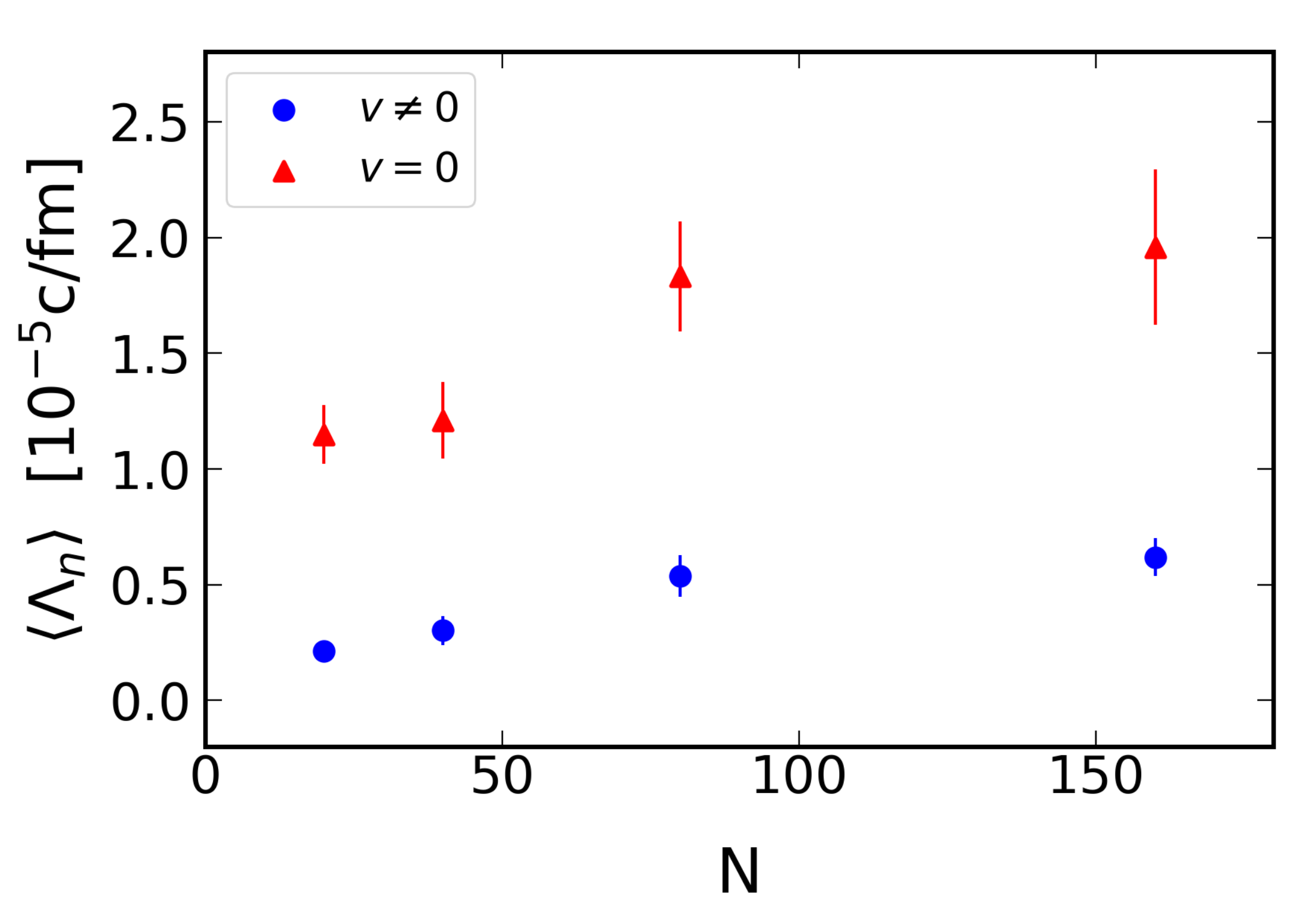}
\caption{Average value of $\left< \Lambda_n \right>$ within the time interval $t=0-1000$ fm/c and for neutron clumps with $N=20$, 40, 80 and 160. The error bars include the statistical deviation of the calculations from their average values. The filled diamonds (circles) represent the solution of Eqs. \eqref{hfbtd1}-\eqref{hfbtd2} with the particle-particle interaction set to zero, $v(x,x')=0$ ($v(x,x')\neq 0$). }
\label{diffus3}
\end{center}
\end{figure}

Another way to assess the relevance of the two-body interaction on the tunneling probability is to set $v(x,x')=0$ at $t>0$. The result is displayed in Figure \ref{diffus3} where the average value of $\left< \Lambda_n \right>$ is shown within the time interval $t=0-1000$ fm/c. In the figure, the error bars take into account the statistical deviation from the mean values. The filled diamonds (circles) represent the calculations using Eqs. \eqref{hfbtd1}-\eqref{hfbtd2} with the particle-particle interaction set to zero, i.e., $v(x,x')=0$ ($v(x,x')\neq 0$). One notices that the tunneling probabilities increase by about $\simeq 4-5$ times when the neutron-neutron interaction is turned off, as claimed above. It is clear that this effect is not treatable within the Gamow model and requires a microscopic formulation of the process when residual nucleon-nucleon interactions play a role.

It is well known that pairing correlations play an important role in nuclear reactions involving two-neutron transfer \cite{bes/1969, Broglia1968}. The effect tunneling enhancement due to pairing becomes evident in transparent analytical models for, e.g.,  the transfer of Cooper pairs and the effects of particle compositeness, as was proved in Ref. \cite{Flambaum2005,BertulaniJPG2007}. In our microscopic model it is easy to assess the role of pairing by turning on and off the pairing density matrix $\kappa \left(x, x^{\prime}\right)=0$ in the TDHFB equations \eqref{hfbtd1}-\eqref{hfbtd2}. When it is off, the method is equivalent to solving the much simpler TD-Hartree-Fock equations. Intriguingly, we have found that the $\Lambda_n$ rates are not strongly modified by paring, with the transfer rates changing by less than 3\% for $N=160$ and 5\% for $N=20$. In fact, our 1D model shows that there is a dominance of single neutron transfer by tunneling. It is also worthwhile mentioning that there is no one-to-one correlation of the process we study here with that involved in neutron transfer in nucleus-nucleus collisions. In the later case, two time-scales are of relevance: (a) One is the reaction time; (b) Another is the neutron-pair tunneling time. The effect we consider in this work only involves one timescale in which the neutrons flow to the less populated regions of space. Our results show that the effect of the pairing energy is unlikely to influence the total energy rearrangement of the system leading to a suppression of single-neutron transfer and favoring  the transfer of a neutron pair.

Finally, we look at the variation of the transfer rate as a function of the distance between the
neutron clumps. The same physics properties obtained for $D=20$ fm and discussed above also
emerges with increasing values of $D$. Only the computation time increases exponentially due to the
exponential decrease of the tunneling probabilities with the  separation distance. This
property is visible in Table \ref{table2} where the aforementioned parameters for the initially
prepared states are used. The diffusion rates resulting from TDHFB calculations are about a factor 6-12 smaller than those predicted by the Gamow model.

\begin{table}[h]
\begin{centering}
\begin{tabular}{|c|c|c|c|c|c|c|c|c|c|c|}
\hline
D & 20 fm& 35 fm & 50 fm \\
\hline
$\left< \Lambda_n \right>$ [c/fm] & $5.86\times 10^{-4}$ & $1.13\times 10^{-5}$& $0.334\times 10^{-5}$\\
$\Lambda_G$ [c/fm] & $ 5.70 \times 10^{-3}$& $8.27\times 10^{-4}$& $1.95\times 10^{-4}$	\\
\hline
\end{tabular}
\caption{Average tunneling probabilities $\left< \Lambda_n \right>$ dependence on the separation
  distance for $N=160$ neutron bubbles with the aforementioned parameters used for the initial many-body wave-function. Also shown are WKB results calculated with Eq. \eqref{Gamow}. \label{table2}}
  \end{centering}
\end{table}

\section{FRBs and magnetars}

Fast radio bursts (FRBs) form a recently discovered astrophysical electromagnetic phenomenon. They have an unknown origin, being bright and of brief duration, with typical time duration within
a range from $\sim30~\mu$s to $\sim20~$ms~\cite{gajjar/2018, michilli/2018, katz/2018}. It is possible that the emission process involves relativistic electron-positron beams with large Lorentz factors $\gamma$, similar to a pulsar radiation mechanism. Electron-position pairs, $e^{\pm}$, are created and  further accelerated
to ultra-relativistic speeds around the polar cap region of the star. A radiation coherence follows leading $N$ particles to radiate with a $N^2$ strength, by far larger than
single-particle emission~\cite{cordes/2019}. Magnetars (SGRs/AXPs) also manifest a pulsar-like
mechanism, with the
radiated power believed to arise from huge magnetic fields ($B\sim10^{12}-10^{15}$~G) instead
of star rotation. In other words, the decay of the hyper-strong magnetic fields is responsible for the cataclysmic emission. Unpredictable and unknown instabilities occurring within these sources are possibly  responsible for the bursting/flaring activities producing  X- and gamma-ray bursts lasting
few milliseconds to tens of seconds. Recently, it was raised the possibility that FRBs originate from
magnetars~\cite{margalit/2018}. We augment these ideas by proposing that the trigger mechanism for the short X/gamma-ray bursts occurring in magnetars and leading to FRBs are due to the diffusion of neutrons in a transiently created inhomogeneous density environment within the star crust. As we have shown, the neutron diffusion can occur in a short time scale within regions of the crust where neutron-rich nuclei are formed by a precursor phenomenon such as a star quake. The diffusion almost instantaneously promotes neutrons to flow to the neighborhoods poor in neutron content, yielding beta and gamma decays which further interact with the strong magnetic fields causing the flares or bursts. The relativistic particles are produced in regions where the density is larger than
$10^6~\mathrm{g/cm^3}$ and will move freely with kinetic energies in the MeV range. The particles generated are responsible for the instabilities in the region, perturbing
the magnetic field, which will
lead to a surge in electromagnetic emission. These particles in the crust and inner magnetosphere will further promote a $e^{\pm}$ production cascade,
yielding a high-energy radiation due to synchrotron radiation and inverse Compton
scattering. Moreover, such a mechanism can also drive Alfvén waves and a large-amplitude wave packet is 
launched due to the disturbance near the surface of the magnetar \cite{kumar/2020}. Again, part of the wave energy is converted into coherent radio emission within a region of few tens of the neutron star radii.

We now provide a rough estimate of amount of energy that our proposed mechanism can release in a burst, $E_{\rm burst}$. We assume that the average separation energy of the less bound neutrons is of the order $\epsilon \sim 1$ MeV and that the subsequent  tunneling of neutrons to a neighboring site will release the same amount of energy. Therefore, the burst can be
estimated as
\begin{equation}\label{burst}
E_{\rm burst} \sim \epsilon N_{\rm imp}
\Lambda_{\rm G} \Delta t \sim \frac{\epsilon N_{\rm
    imp} 10^2L}{\alpha D} \sqrt{\frac{2U_{0}}{mc^2}}\exp\left(-{2D\sqrt{2mc^2\epsilon} \over \hbar c} \right),
\end{equation}
where, for simplicity, we take  the tunneling rate $\Lambda_{\rm G}$ from Eq. \eqref{Gamow}. This procedure neglects the many-body effects we have considered so far, but can be modified to account for such effects in a more complicated calculation. Because the energy released by the tunneling process occurs almost instantaneously, the time during which the explosive emission occurs  is the same as the time  during which the wave sweeps through the crust changing its local densities, i.e., $\Delta t\approx L/v
= 10^{2}L/c$, where $L\approx 2$ km. In this equation $N_{\rm imp} $ is the number of impurities (neutron bubbles, or clumps) coherently participating in the process. The bubbles are assumed to
have a dimension $\alpha D$, with $\alpha > 1$ adjusted so that the bubbles sizes can be varied at will.

\begin{figure}[!h]
\begin{center}
\includegraphics[scale=0.7]{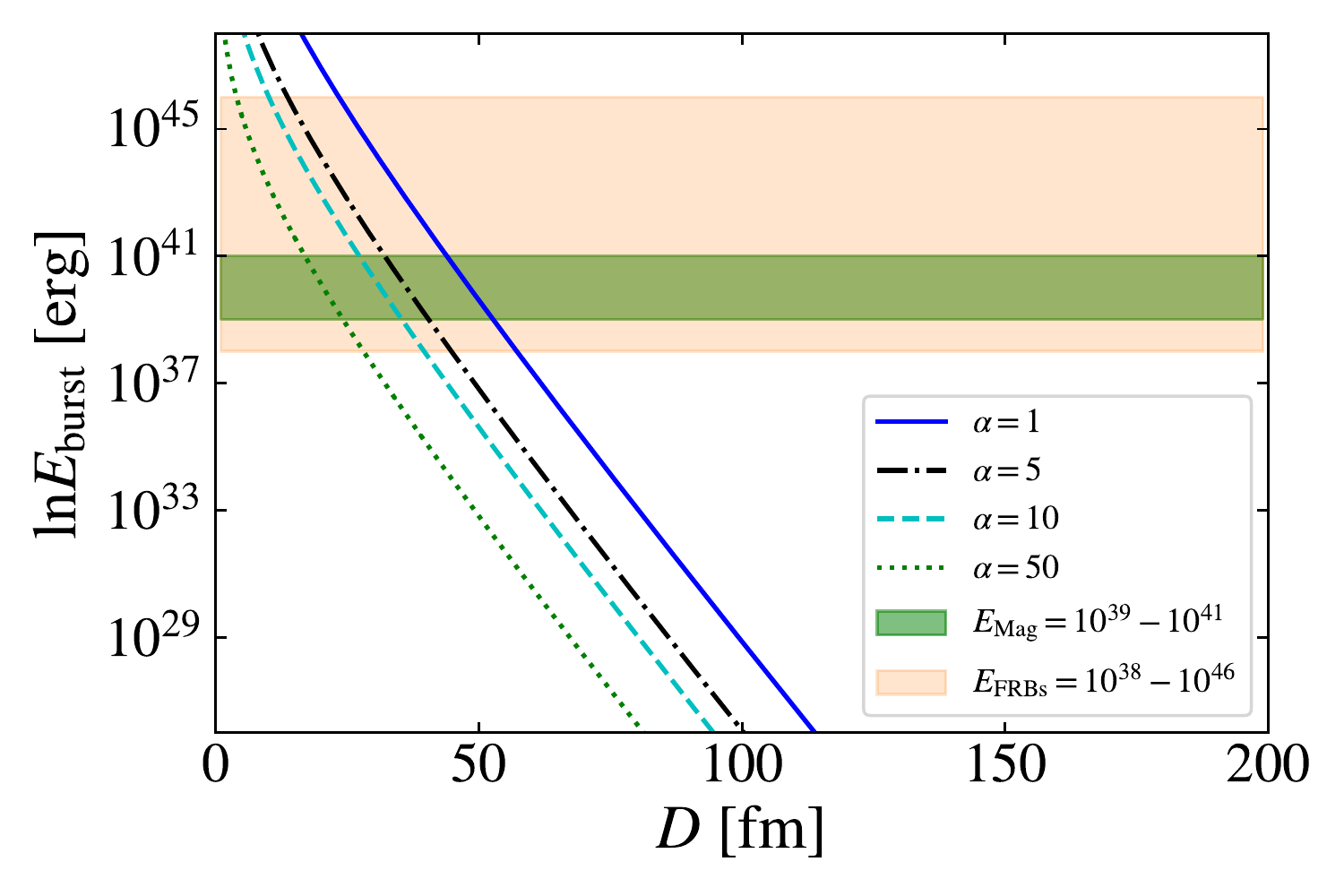}
\caption{Energy released in the form of bursts as function of the separation distance $D$ and neutron dense sites with sizes determined by $\alpha=1, 5, 10, 50$. The
  green horizontal band limits the observed values of short bursts from magnetars, while
  the orange band represents the limits of the observed values for FRBs.}
\label{energy}
\end{center}
\end{figure}

To determine the number of neutron-rich clusters participating in the coherent mechanism, we consider the polar cap region within a twisted magnetic field configuration with a radius $R_{\rm p} =
R\sin\theta_{\rm p}= R\sqrt{(R\Omega/c)^n/(15+17n)/32}$, with $n$ evolving from $n<1$ to $n=1$
~\cite{tong/2019}. We estimate the height under the polar cap, $h_{\rho_0}$, starting from
the star surface and ending in a layer with density $\rho=\rho_0$ ($\rho_0=0.17$ fm$^{-3}$ being the nuclear matter saturation density). This yields $h_{\rho_0}\approx 2$~km. Therefore, we obtain the volume of interest, $V = \pi R_{\rm p}^2h_{\rho_o}$, comprising the emission region within magnetars. The number of neutron clumps in this region follows as
  \begin{equation}
  N_{\rm imp} = {V\over 4/3 \pi (\alpha D)^3}\sim {3R^2 h_{\rho_o} \over 4\alpha^3 D^3}\frac{(R\Omega/c)^n}{(15+17n)/32}.
\end{equation}
We study four possible scenarios with $\alpha = 1$, 5, 10 and 50, allowing for 
neutron inhomogeneities which can also be much larger in size than their separation distances. In Figure~\ref{energy} we plot
the energy emitted during such a burst mechanism as a function of the separation distance $D$ between the neutron clumps. We assume a polar cap
radius $R_{\rm p} = 2$ km. The green shaded horizontal band accounts for the values of short bursts observed in magnetars $\sim 10^{39}-10^{41}~{\rm
  erg/s}$~\cite{turolla/2015} while the shaded orange band represent the energy burst values observed for FRBs $\sim 10^{38}-10^{46}~{\rm
  erg/s}$~\cite{zhang/2020a}.

It is noticeable from Figure \ref{energy} that our proposed mechanism for the aforementioned bursts are comparable to the astronomically observed ones for magnetars. The same applies to the observed values of FRBs, assuming that the dense neutron sites are separated by 
distances smaller than $D\sim 50$ fm.  Figure \ref{energy} also shows that smaller size impurities favor larger energy yields. Evidently, these are crude estimates, relying on the WKB approximation. Corrections due to the microscopic behavior of the many-body neutron systems and further medium interactions can easily change these figures by an order of magnitude. A microscopic calculation carried out in a three-dimensional lattice can also increase the tunneling rates because the increased number of neighbors can offer more diffusion/tunneling opportunities. Many features displayed as an outcome of our one-dimensional  model are likely to remain valid in a more complicated three-body lattice dynamical modeling.

\section{Conclusions}

 In this work (see also Ref. \cite{bertulani/2021a}), we have reported a microscopic TDHFB calculation for the neutron diffusion rates in transient neutron stars due to neutron diffusion/tunneling
 from neutron-rich to neutron-poor regions. We considered neutron clumps consisting of 20, 40, 80 and 160 neutrons taking into account that nuclear barriers will form between the impurities and the surroundings. Our model is one dimensional in nature, as are other models used to infer the gross properties of many-body interacting systems, such as the well-known Ising model. Our model allows the understanding of key features of the time evolution of strongly interacting particles.

 Our main findings in this pioneer study include, albeit they are not limited to,
 \begin{itemize}
 \item there are clear differences between calculations based on perturbation theory such as the WKB model and a time-dependent microscopic modeling. The detachment  and diffusion of neutrons within an inhomogeneous neutron environment is poorly inferred from a perturbative treatment,
 \item the role of neutron pairing is subtle and strongly depends on the size and density of the system under scrutiny. We did not observe an enhancement of the tunneling rates for a neutron pair compared to single neutron transfer, as has been usually advertised in heavy-ion transfer reactions,
 \item neutron diffusion by tunneling cannot be ruled out as a possible mechanism for the occurrence of electromagnetic bursts from magnetars.  Our estimates have shown that such a mechanism is possible.
 \end{itemize}

 Our studies conclude that the subject of dynamical density homogenization and
 isospin diffusion in stellar environments and in nuclear reactions is far from being understood and therefore deserves a more extensive theoretical study.
 Perturbative estimates such as those based on WKB transmission probabilities are likely to yield
 poor results, due to the dynamical rearrangement of strongly interacting systems. There are other subtleties involved, 
 such as the dynamical contribution of the mean field and the particle-particle interaction to the time evolution of the local energies within the system. The microscopic approach is
 rich in details emerging for the transmission and diffusion rates.
 
 This work also provides further insights on the role of the microscopic behavior of tiny nucleon inhomogeneities within a macroscopic object such as a neutron star and its dynamical evolution following a disruptive process and the ensuing explosive phenomena. The time evolution of density inhomogeneities and their energy release by means of sudden nuclear diffusion processes are also of relevance for the dynamical description of neutron star mergers. 

 \section*{Acknowledgements}
This work has been supported in part by the U.S. DOE Grant No. DE-FG02-08ER41533. We would like to thank the Brazilian funding agency CNPq and the INCT-FNA, Brazil, for the financial support.

\section*{References}

\begin{thebibliography}{10}
\expandafter\ifx\csname url\endcsname\relax
  \def\url#1{{\tt #1}}\fi
\expandafter\ifx\csname urlprefix\endcsname\relax\def\urlprefix{URL }\fi
\providecommand{\eprint}[2][]{\url{#2}}

\bibitem{TsangPRL.92.062701}
Tsang M~B, Liu T~X, Shi L, Danielewicz P, Gelbke C~K, Liu X~D, Lynch W~G, Tan
  W~P, Verde G, Wagner A, Xu H~S, Friedman W~A, Beaulieu L, Davin B, de~Souza
  R~T, Larochelle Y, Lefort T, Yanez R, Viola V~E, Charity R~J and Sobotka L~G
  2004 {\em Phys. Rev. Lett.\/} {\bf 92}(6) 062701

\bibitem{ShiPRC.68.064604}
Shi L and Danielewicz P 2003 {\em Phys. Rev. C\/} {\bf 68}(6) 064604

\bibitem{Chugunov2018}
Chugunov A~I 2018 {\em Monthly Notices of the Royal Astronomical Society:
  Letters\/} {\bf 483} L47--L51 ISSN 1745-3925

\bibitem{fryer/2006}
Fryer C~L, Herwig F, Hungerford A and Timmes F~X 2006 {\em The Astrophysical
  Journal\/} {\bf 646} L131--L134

\bibitem{piro/2011}
Piro A~L and Ott C~D 2011 {\em The Astrophysical Journal\/} {\bf 736} 108

\bibitem{metzger/2018a}
Metzger B~D, Beniamini P and Giannios D 2018 {\em The Astrophysical Journal\/}
  {\bf 857} 95

\bibitem{bertulani/2021a}
Bertulani C~A and Lobato R~V 2021 {\em The Astrophysical Journal\/} {\bf 912}
  105

\bibitem{ringschuck80}
Ring P and Schuck P 1980 {\em The nuclear many-body problem\/} (New York:
  Springer-Verlag)

\bibitem{Gamow1928}
Gamow G 1928 {\em Zeitschrift f{\"u}r Physik\/} {\bf 51} 204--212

\bibitem{bes/1969}
Bes D~R and Sorensen R~A 1969 The {{Pairing-Plus-Quadrupole Model}} {\em
  Advances in {{Nuclear Physics}}: {{Volume}} 2\/} ed Baranger M and Vogt E
  ({New York, NY}: {Springer US}) pp 129--222 ISBN 978-1-4684-8343-7

\bibitem{Broglia1968}
Broglia R~A, Riedel C and Soerensen B 1968 {\em Nucl. Phys. A\/} {\bf 107}

\bibitem{Flambaum2005}
Flambaum V~V and Zelevinsky V~G 2005 {\em Journal of Physics G: Nuclear and
  Particle Physics\/} {\bf 31} 355--360

\bibitem{BertulaniJPG2007}
Bertulani C~A, Flambaum V~V and Zelevinsky V~G 2007 {\em Journal of Physics G:
  Nuclear and Particle Physics\/} {\bf 34} 2289--2295

\bibitem{gajjar/2018}
Gajjar V, Siemion A~P~V, Price D~C, Law C~J, Michilli D, Hessels J~W~T,
  Chatterjee S, Archibald A~M, Bower G~C, Brinkman C, {Burke-Spolaor} S, Cordes
  J~M, Croft S, Enriquez J~E, Foster G, Gizani N, Hellbourg G, Isaacson H,
  Kaspi V~M, Lazio T~J~W, Lebofsky M, Lynch R~S, MacMahon D, McLaughlin M~A,
  Ransom S~M, Scholz P, Seymour A, Spitler L~G, Tendulkar S~P, Werthimer D and
  Zhang Y~G 2018 {\em The Astrophysical Journal\/} {\bf 863} 2

\bibitem{michilli/2018}
Michilli D, Seymour A, Hessels J~W~T, Spitler L~G, Gajjar V, Archibald A~M,
  Bower G~C, Chatterjee S, Cordes J~M, Gourdji K, Heald G~H, Kaspi V~M, Law
  C~J, Sobey C, Adams E~a~K, Bassa C~G, Bogdanov S, Brinkman C, Demorest P,
  Fernandez F, Hellbourg G, Lazio T~J~W, Lynch R~S, Maddox N, Marcote B,
  McLaughlin M~A, Paragi Z, Ransom S~M, Scholz P, Siemion A~P~V, Tendulkar S~P,
  Van~Rooy P, Wharton R~S and Whitlow D 2018 {\em Nature\/} {\bf 553} 182--185

\bibitem{katz/2018}
Katz J~I 2018 {\em Progress in Particle and Nuclear Physics\/} {\bf 103} 1--18

\bibitem{cordes/2019}
Cordes J~M and Chatterjee S 2019 {\em Annual Review of Astronomy and
  Astrophysics\/} {\bf 57} 417--465

\bibitem{margalit/2018}
Margalit B and Metzger B~D 2018 {\em The Astrophysical Journal\/} {\bf 868} L4

\bibitem{kumar/2020}
Kumar P and Bo{\v s}njak {\v Z} 2020 {\em Monthly Notices of the Royal
  Astronomical Society\/} {\bf 494} 2385--2395

\bibitem{tong/2019}
Tong H 2019 {\em Monthly Notices of the Royal Astronomical Society\/} {\bf 489}
  3769--3777

\bibitem{turolla/2015}
Turolla R, Zane S and Watts A~L 2015 {\em Reports on Progress in Physics\/}
  {\bf 78} 116901

\bibitem{zhang/2020a}
Zhang B 2020 {\em Nature\/} {\bf 587} 45--53

\end{thebibliography}

\end{document}